\documentclass[onecolumn,manuscript]{revtex4}
\usepackage{amssymb}

\usepackage{epsfig}
\usepackage[english]{babel}
\usepackage{latexsym}
\usepackage{graphics}
\usepackage{subfigure}
\usepackage{epsfig}
\usepackage{graphicx}
\usepackage{dcolumn}
\usepackage{amsmath}

\def\PRA{{Phys.~Rev.~A} }

\def\JPB{{J.~Phys.~B: At. Mol. Opt. Phys.} }
\def\CPB{{Chin.~Phys.~B} }
\def\PRL{{Phys.~Rev.~Lett.} }

\begin{document}

\preprint{AUR1028}

\title{Nonsequential double ionization of helium in IR+XUV two-color laser fields II$:$ Collision-excitation ionization process}

\author{Facheng Jin$^{1,2}$, Jing Chen$^{3}$,  Yujun Yang$^{4}$,  Xiaojun Liu$^{5}$, Zong-Chao Yan$^{6}$, and Bingbing Wang$^{1}$}
\email{}

\address{$^1$Laboratory of Optical Physics, Beijing National Laboratory for Condensed Matter Physics, Institute of Physics, Chinese Academy of Sciences, Beijing 100190, China}
\address{$^2$College of Science, Xi'an Aeronautical University, Xi'an 710077,  China}
\address{$^{3}$Institute of Applied Physics and Computational Mathematics, P. O. Box 8009, Beijing 100088, China}
\address{$^4$Institute of Atomic and Molecular Physics, Jilin University, Changchun 130012, China}
\address{$^5$State Key Laboratory of Magnetic Resonance and Atomic and Molecular Physics, Wuhan Institute of Physics and Mathematics, Chinese Academy of Sciences, Wuhan 430071, China}
\address{$^6$Department of Physics, University of New Brunswick,Fredericton, New Brunswick, Canada E3B 5A3}
\date{\today}

\begin{abstract}
 The collision-ionization mechanism of nonsequential double ionization
 (NSDI) process in IR+XUV two-color laser fields [\PRA \textbf{93}, 043417
 (2016)] has been investigated by us recently. Here we extend this work to
 study the collision-excitation-ionization (CEI) mechanism of NSDI
 processes in the two-color laser fields with different laser conditions.
 It is found that the CEI mechanism makes a dominant contribution to the
 NSDI as the XUV photon energy is smaller than the ionization threshold of
 the He
 $^+$ ion, and the momentum spectrum shows complex interference
 patterns and symmetrical structures. By channel analysis, we find that,
 as the energy carried by the recollision electron is not enough to excite
 the bound electron, the bound electron will absorb XUV photons during
 their collision; as a result, both forward and backward collisions make a
 comparable contributions to the NSDI processes. However, it is found
 that, as the energy carried by the recollision electron is large enough
 to excite the bound electron, the bound electron does not absorb any XUV
 photon and it is excited only by sharing the energy carried by the
 recollsion electron, hence the forward collision plays a dominant role on
 the NSDI processes. Moreover, we find that the interference patterns of
 the NSDI spectra can be reconstructed by the spectra of two
 above-threshold ionization (ATI) processes, which may be used to analyze
 the structure of the two separate ATI spectra by NSDI processes.

\end{abstract}

\par\noindent

\maketitle

\par\noindent
$^*$ wbb@aphy.iphy.ac.cn
\section{INTRODUCTION}
Nonsequential double ionization (NSDI) of atoms and molecules in a strong
infrared (IR) laser field has continued to attract considerable
interest~\cite{Becker2012} since the first observation of the knee structure
in the doubly charged ion yield curve as a function of laser intensity for
Xe atom~\cite{Huillier1983}. Up to now, it is widely accepted that the
recollision picture is responsible for the NSDI~\cite{Corkum1993}. In this
picture, one electron is released via a tunneling process, then accelerated,
and finally driven back to collide with another electron making two
electrons simultaneously ionized by the IR laser field, which is called the
collision-ionization (CI) process; whereas if the second electron is pumped
to an excited state, and then ionized subsequently by the IR laser field,
this process is called the collision-excitation-ionization (CEI). With great
advances in the study of the NSDI by many theoretical
approaches~\cite{Li2009,Hao2014,Parker2006,Chen,Ye2008,Ho2006,Mauger2010,Goreslavski2001,Goreslavskii2001,Kopold2000,Faria2005,Wang2012,Jin1,Jin2},
for example, solving the full-dimensional time-dependent Schr\"{o}dinger
equation~\cite{Parker2006}, the S-matrix
theory~\cite{Goreslavskii2001,Kopold2000,Faria2005,Hao2014,Goreslavski2001},
semiclassical~\cite{Chen,Ye2008,Li2009} and
classical~\cite{Ho2006,Mauger2010} models as well as the frequency-domain
theory~\cite{Wang2012,Jin1,Jin2}, both the CI and CEI mechanisms have been
investigated in the NSDI process. Especially in recent years, more attention
has been paid to the CEI mechanism of NSDI because it involves more complex
electron correlation dynamics and  may give rise to nontrivial NSDI
features. Indeed, a number of interesting NSDI phenomena have been observed
recently~\cite{Bergues2012,Huang2013,Zhang2015,Shaaran,Maxwell2016,Hao2014,Sun2014,Dong2016,Yuan2016,Eckart2016},
such as the cross-shaped in the correlated electron momentum distribution
from NSDI with a near-single-cycle~\cite{Bergues2012,Huang2013}, the
dependence of momentum distribution on laser intensities and
carrier-envelope phases~\cite{Shaaran,Maxwell2016}, and quantum and
depletion effects below the threshold intensity~\cite{Hao2014}, as well as
the nonstructured momentum distribution of high-Z rare-gas
atoms~\cite{Sun2014}. These results are believed to be closely related to
the CEI mechanism and further stimulate the study of NSDI process caused by
CEI.

With rapid development of free-electron laser
technology~\cite{Rosenblum2016,Shintake2008,Emma2010} and application of the
high-order harmonic generation~\cite{Popmintchev2010,Chini2014,Pan2013},
NSDI can take place in an extreme ultraviolet (XUV) laser field. Applying
XUV light source, a great progress has been made in understanding the NSDI
dynamics in this short wavelength regime~\cite{Briggs2000,Peng2015}. For
example, the virtual sequential model~\cite{Jiang2015} and the joint angular
distribution~\cite{Zhang2011,Liua2015} show the effect of the electron
correlation in the NSDI process. More recently, NSDI dynamics has been
further explored in the combined IR and XUV fields. Hu~\cite{Hu2013} first
calculated the double ionization of helium in IR+XUV two-color laser fields
by \textit{ab initio} calculation. It was found that the electron
correlation can significantly enhance the NSDI probability by controlling
the XUV laser field to excite the bound electron of He$^{+}$ ion. Liu
\textit{et al.}~\cite{Liu} investigated the dependence of joint
photoelectron angular distributions on the energy sharing of two ionized
electrons by solving the time-dependent Schr\"{o}dinger equation in full
dimensionality. It was found that the joint photoelectron angular
distribution in IR+XUV two-color laser fields is different from that in a
monochromatic XUV laser field.

With the help of formal scattering theory~\cite{Mann}, the frequency-domain
theory based on the nonperturbative quantum electrodynamics was developed by
Guo \textit{et al.}~\cite{Guo1992}. This approach has been successfully
extended to deal with the recollision processes, such as the high-order
above-threshold ionization (HATI)~\cite{Wang2007,Wang2010}, high-order
harmonic generation (HHG)~\cite{Gao2000,Fu2001} and the NSDI~\cite{Wang2012}
in a monochromatic IR laser field. One major advantage of this
frequency-domain theory is that all of the recollision processes can  be
decoupled into certain channels,  which can be further investigated
separately.  Also, the approach can save a large amount of computation time
because of its nature of time-independence. Recently, the frequency-domain
theory has been employed to investigate the above-threshold ionization (ATI)
in IR+XUV two-color laser fields~\cite{Zhang2013,Liu2015}. Furthermore, we
have studied the momentum spectrum of two ionized electrons and analyzed the
formation of momentum spectrum in the NSDI due to the CI mechanism in the
two-color laser fields~\cite{Jin1,Jin2}. In this paper, we shall extend the
frequency-domain theory to investigate the CEI mechanism of NSDI in IR+XUV
two-color laser fields. By the channel analysis, we investigate the channel
contributions to the momentum distribution of two ionized electrons as well
as the contributions of the forward and backward collisions on NSDI.
Moreover, by employing the saddle-point approximation, we obtain the energy
conservation equations satisfied by the two ionized electrons and illustrate
the formation of the momentum spectrum.

This paper is organized as follows. Section II introduces the theoretical
method. In Sec. III, we present the  numerical results and a comprehensive
analysis of them. Finally, in Sec. IV our conclusions are given. Atomic
units are used throughout unless otherwise stated.

\section{THEORETICAL METHOD}

The Hamiltonian for the system of a two-electron atom in two-color linearly
polarized laser fields with frequencies of $\omega_1$ and $\omega_2$  is
\cite{Guo1992}
\begin{equation}\label{1}
H = {H_0} +U + V,
\end{equation}
where
\begin{equation}\label{2}
{H_0} = \frac{{( - i\nabla_1 )}^2}{2} +\frac{{( - i\nabla_2 )}^2}{2}+ {\omega_1}{N_{1}} + {\omega_2}{N_{2}}
\end{equation}
is the free electron and photon energy operator for the laser-matter system.
$N_{i}$ is the photon number operator for laser field with frequency
$\omega_i$ for $i=1,2$. The Coulomb potential is $U=U_1+U_2+U_{12}$, where
$U_i$ for $i=1,2$ is the interaction potential between electron and nuclei,
and $U_{12}$ is the interaction potential between the two electrons.
Finally, the electron-photon interaction potential is $V=V_1+V_2$, where
\begin{equation}\label{3}
\begin{array}{l}
V_j=-[(-i\nabla_j)\cdot \textbf{A}_1(-\textbf{k}_1\cdot \textbf{r}_j)+(-i\nabla_j)\cdot \textbf{A}_2(-\textbf{k}_2\cdot \textbf{r}_j)] \\
 ~~~~~~~ +\frac{1}{2}[\textbf{A}_1(-\textbf{k}_1\cdot \textbf{r}_j)+\textbf{A}_2(-\textbf{k}_2\cdot \textbf{r}_j)]^2,\\
\end{array}
\end{equation}
$\textbf{A}_{s}(-\textbf{k}_{s}\cdot \textbf{r}_j)=(2V_{s} \omega_{s})^
{-1/2} ({\hat{\varepsilon}}_{s} e^{i \textbf{k}_{s}\cdot \textbf{r}_j} a_{s}
+c_.c_.)$ is the vector potential with $j=1,2$, $\textbf{k}_{s}$ is the wave
vector, $\hat{\varepsilon}_{s}$ is the polarization vector,
$a_{s}(a^{\dag}_{s})$ is the annihilation (creation) operator, and $V_{s}$
is the normalization volume of the laser field with the frequency
$\omega_{s}$ for $s=1,2$.

The transition matrix from the initial state $|\Psi_i\rangle$ to the final
state $|\Psi_f\rangle$ for a NSDI process caused by the CEI mechanism is
\cite{Wang2012}
\begin{equation}\label{4}
{T_{\rm{CEI}}} =\langle \Psi_f|V_2\frac{1}{E_f-H-i\varepsilon}U_{12}\frac{1}{E_i-H+i\varepsilon}V_1|\Psi_i\rangle.
\end{equation}
In the above, the initial state for the system is $|\Psi_i\rangle=|\Psi_{i
n_1
n_2}(\textbf{r}_1,\textbf{r}_2)\rangle=|\Phi_1(\textbf{r}_1,\textbf{r}_2)\rangle
\otimes|n_1\rangle \otimes |n_2\rangle$, where
$|\Phi_1(\textbf{r}_1,\textbf{r}_2)\rangle$ is the ground state of the atom,
and $|n_j\rangle$ is the Fock state of the laser field with the photon
number $n_j$ for $j=1,2$. The initial energy of the system is $E_i =  -
(I_{p_1}+I_{p_2}) + ({n_1} + 1/2){\omega_1} + ({n_2} + 1/2){\omega _2}$ with
$I_{p_1} (I_{p_2})$ being the first (second) ionization potential of the
atom. The final state $|\Psi_f\rangle =|\Psi_{\textbf{p}_1 \textbf{p}_2 k_1
k_2}(\textbf{r}_1,\textbf{r}_2)\rangle$ is the quantized-field Volkov state
of two electrons in a two-color laser field~\cite{Guo1992,Wang2012}
\begin{equation}\label{5}
\begin{array}{l}
{|\Psi_{\textbf{p}_1 \textbf{p}_2 k_1 k_2}(\textbf{r}_1,\textbf{r}_2)\rangle} = V_e^{ - 1} \exp\{i[(\textbf{p}_1+u_{p_1}\textbf{k}_1+u_{p_2}\textbf{k}_2)\cdot \textbf{r}_1 +(\textbf{p}_2+u_{p_1}\textbf{k}_1+u_{p_2}\textbf{k}_2)\cdot \textbf{r}_2]\}\\
  ~~~~~~~~~~~~~~\times \sum\limits_{{j_1} =  - {k_1},{j_2} =  - {k_2}}^\infty \aleph_{j_1 j_2}(\zeta_1)^{*}\exp \{-i[j_1 (\textbf{k}_1 \cdot \textbf{r}_1 +\phi_1)+j_2(\textbf{k}_2 \cdot \textbf{r}_1 +\phi_2)]\} \\
  ~~~~~~~~~~~~~~   \times \sum\limits_{{j_3} =  - {k_1}-{j_1},{j_4} =  - {k_2}-{j_2}}^\infty \aleph_{j_3 j_4}(\zeta_2)^{*}\exp \{-i[j_3 (\textbf{k}_1 \cdot \textbf{r}_2 +\phi_1)+j_4(\textbf{k}_2 \cdot \textbf{r}_2 +\phi_2)]\} \\
 ~~~~~~~~~~~~~~   \times |k_1+j_1+j_3,k_2+j_2+j_4\rangle,
 \end{array}
\end{equation}
with the energy $E_f ={\textbf{p}_1}^2/2+{\textbf{p}_2}^2/2 + ({k_1} +
1/2){\omega_1} + ({k_2} + 1/2){\omega _2}+2u_{p_1}\omega_1
+2u_{p_2}\omega_2$, where $u_{p_j}$ and $\phi_j$ are the ponderomotive
energy in units of frequency $\omega_j$ and the initial phase of the laser
field with frequency $\omega_j$ for $j=1,2$, respectively. Under the strong
field approximation, the interaction between the two ionized electrons is
ignored by assuming that the two electrons are separated far away from each
other in the final state Eq.~(\ref{5}). The possible effects of Coulomb interaction of the two electron in the final states may reduce the density distribution in the low momentum region on the momentum spectra. The term $\aleph_{\gamma_1 \gamma_2}
(\zeta _i)$ for $i=1,2$ in Eq.~(\ref{5}) is the generalized Bessel function,
which can be written as \cite{Guo1992}
\begin{equation}\label{6}
\begin{array}{l}
 {\aleph_{{\gamma_1} {\gamma_2}}}(\zeta_i) = \sum\limits_{{\gamma_3}{\gamma_4}{\gamma_5}{\gamma_6}} {{J_{ - {\gamma_1} + 2{\gamma_3} + {\gamma_5} + {\gamma_6}}}} ({\varsigma_1}){J_{ - {\gamma_2} + 2{\gamma_4} + {\gamma_5} - {\gamma_6}}}({\varsigma_2}) \\
 ~~~~~~~~~~~~~~ \times {J_{ - {\gamma_3}}}({\varsigma_3}){J_{ - {\gamma_4}}}({\varsigma_4}){J_{ - {\gamma_5}}}({\varsigma_5}){J_{ - {\gamma_6}}}({\varsigma_6}), \\
 \end{array}
\end{equation} 	
where
\begin{equation}\label{7}
\begin{gathered}
  {\varsigma_1} = 2\sqrt {\frac{{{u_{p_1}}}}{{{\omega_1}}}} {{\mathbf{\text{\textbf{p}}}}_i}\cdot {{\widehat{{\mathbf{\epsilon}}}}_1},~~~~~~~ {\varsigma_2} = 2\sqrt{\frac{{{u_{p_2}}}}
{{{\omega _2}}}} {{\text{\textbf{p}}}_i} \cdot {{\widehat{{\mathbf{\epsilon}}}}_2}, \hfill \\
{\varsigma_3} = \frac{1}{2}u_{p_1},~~~~~~~~~~~~~~~~~~~ {\varsigma_4} = \frac{1}{2}u_{p_2}, \hfill \\
{\varsigma_5} = 2\frac{{\sqrt {{u_{p_1}}{\omega _1}{u_{p_2}}{\omega _2}} }}
{{{\omega _1} + {\omega _2}}},~~~~~~ {\varsigma_6} = 2\frac{{\sqrt {{u_{p_1}}{\omega _1}{u_{p_2}}{\omega _2}} }}{{{\omega _1} - {\omega _2}}}, \hfill \\
\end{gathered}
\end{equation}	
and $J_m (t)$ is the Bessel function of order $m$. In above expression,
$\textbf{p}_i$ is the momentum of the ionized electron, and
${{\widehat{{\mathbf{\epsilon}}}}_1}$
(${{\widehat{{\mathbf{\epsilon}}}}_2}$) is the polarization direction of the
laser's electric field with frequency $\omega_1$ ($\omega_2$).

In the CEI process, two sets of intermediate states of the atom-laser system
are assumed. One state is that an electron is ionized in a Volkov state and
the other electron is bound in the ground state of $\textmd{He}^+$ ion,
which can be expressed as $|\Psi_{\textbf{p}'_1 m_1
m_2}(\textbf{r}_1)\rangle\otimes |\Phi_1 (\textbf{r}_2)\rangle$
corresponding to the energy $E_m ={\textbf{p}'_1}^2/2 + ({m_1} +
1/2){\omega_1} + ({m_2} + 1/2){\omega _2}+u_{p_1}\omega_1
+u_{p_2}\omega_2-I_{p_2}$. The other intermediate state is that one electron
is ionized in an Volkov state and the other is at the first excited state of
$\textmd{He}^+$ ion, which can be expressed as $|\Psi_{\textbf{p}''_1 l_1
l_2}(\textbf{r}_1)\rangle\otimes |\Phi_2 (\textbf{r}_2)\rangle$
corresponding to the energy $E_l ={\textbf{p}''_1}^2/2 + ({l_1} +
1/2){\omega_1} + ({l_2} + 1/2){\omega _2}+u_{p_1}\omega_1
+u_{p_2}\omega_2-I_{p_{12}}$ with $I_{p_{12}}$ being the ionization
potential of the first excited sate of $\textmd{He}^+$ ion. In our
calculation, we find that the NSDI probability contributed by the first
excited state is much larger than that by other higher excited states.
Therefore, we only consider the contribution from the first excited state of
He$^{+}$ ion to the NSDI. $|\Psi_{\textbf{p}'_1 m_1
m_2}(\textbf{r}_1)\rangle$ and $|\Psi_{\textbf{p}''_1 l_1
l_2}(\textbf{r}_1)\rangle$ are the well-known Volkov states, which can be
written as \cite{Guo1992}
\begin{equation}\label{8}
\begin{array}{l}
 {|\Psi_{\textbf{p}_1 k_1 k_2}(\textbf{r}_1)\rangle} = V_e^{ - 1/2}\sum\limits_{{j_1} =  - {k_1},{j_2} =  - {k_2}}^\infty \exp \{i[\textbf{p}_1+(u_{p_1}-j_1)\textbf{k}_1+(u_{p_2}-j_2)\textbf{k}_2]\cdot\textbf{r}_1\} \\
 ~~~~~~~~~~~~~~~ \times \aleph_{j_1 j_2}(\zeta_1)^{*} \exp[-i(j_1\phi_1+j_2 \phi_2)]|k_1+j_1,k_2+j_2\rangle. \\
 \end{array}
\end{equation}
By applying the completeness relation of the intermediate states together with the initial and final states,  Eq.~(\ref{4}) can be rewritten as
\begin{equation}\label{9}
\begin{array}{l}
 {T_{\rm{CEI}}} = \pi^2 \sum\limits_{\textbf{p}''_1 l_1 l_2}^\infty \sum\limits_{\textbf{p}'_1 m_1 m_2}^\infty \langle \Psi_{\textbf{p}_1 \textbf{p}_2 k_1 k_2}(\textbf{r}_1,\textbf{r}_2)|V_2|\Psi_{\textbf{p}''_1 l_1 l_2}(\textbf{r}_1)\Phi_2(\textbf{r}_2)\rangle \langle \Psi_{\textbf{p}''_1 l_1 l_2}(\textbf{r}_1)\Phi_2(\textbf{r}_2)\\
 ~~~~~~~~~~~\times|U_{12}|\Psi_{\textbf{p}_1^{'} m_1 m_2}(\textbf{r}_1)\Phi_1(\textbf{r}_2)\rangle \langle \Psi_{\textbf{p}_1^{'} m_1 m_2}(\textbf{r}_1)\Phi_1(\textbf{r}_2)|V_1 |\Psi_{i n_1 n_2}(\textbf{r}_1,\textbf{r}_2) \rangle \\
 ~~~~~~~~~~~  \times \delta (E_l-E_i)  \delta (E_m-E_i) \\
 ~~~~~~~=\pi^2 \sum\limits_{\textbf{p}''_1 l_1 l_2}^\infty \sum\limits_{\textbf{p}'_1 m_1 m_2}^\infty T_{\rm{ATI2}}T_{\rm{LACE}}T_{\rm{ATI1}}\delta (E_l-E_i)  \delta (E_m-E_i),
 \end{array}
\end{equation}
where
\begin{equation}\label{901}
\begin{array}{l}
 T_{\rm{ATI1}} = \langle \Psi_{\textbf{p}'_1 m_1 m_2}(\textbf{r}_1)\Phi_1(\textbf{r}_2)| V_1 |\Psi_{i n_1 n_2}(\textbf{r}_1,\textbf{r}_2) \rangle \\
~~~~~~~~=V_e^{-1/2}[(u_{p_1}-q_1)\omega_1+(u_{p_2}-q_2)\omega_2]\aleph_{q_1q_2}(\zeta_1^{'})\Phi_1(\textbf{p}_1^{'}),
 \end{array}
\end{equation}
with $q_1=n_1-m_1$ and $q_2=n_2-m_2$ being the photon numbers absorbed by the first electron $e_1$ from the first and second laser fields. The term $T_{\rm{ATI1}}$ represents the ATI process that the first electron $e_1$ is ionized directly from the ground state of $\textmd{He}$ atom to the continuous states, which is called the ATI1 process. Here we make some approximations that the ground state $|\Phi_1(\textbf{r}_1,\textbf{r}_2)\rangle$ of an atom is a production of two hydrogen-like wave functions, and the second electron is in the ground state of He$^+$ ion when the first electron is ionized in the ATI1 process~\cite{Wang2012}. The term $T_{\rm{LACE}}=\langle \Psi_{\textbf{p}''_1 l_1 l_2}(\textbf{r}_1)\Phi_2(\textbf{r}_2)|U_{12}|\Psi_{\textbf{p}_1^{'} m_1 m_2}(\textbf{r}_1)\Phi_1(\textbf{r}_2)\rangle$ represents the laser-assisted collision excitation (LACE) process that the first ionized electron $e_1$ collides with the bound electron $e_2$ and sets it to the first excited state of $\textmd{He}^{+}$ ion. The term $T_{\rm{ATI2}}$ can be rewritten as
 \begin{equation}\label{902}
\begin{array}{l}
 T_{\rm{ATI2}} =V_e^{-1/2}\delta (\textbf{p}_1-\textbf{p}_1^{''})[(u_{p_1}-s_1)\omega_1+(u_{p_2}-s_2)\omega_2]\aleph_{s_1 s_2}(\zeta_2)\Phi_2(\textbf{p}_2),
 \end{array}
\end{equation}
 with $s_1=l_1-k_1$ and $s_2=l_2-k_2$ being the photon numbers absorbed by the second electron $e_2$ from the first and second laser fields. The term $T_{\rm{ATI2}}$ represents the ATI process of the electron $e_2$ from the first excited state of $\textmd{He}^+$ ion to the continuous states, which is called the ATI2 process. By using Eqs.~(\ref{5}) and~(\ref{8}), the transition matrix for the NSDI can be written as
\begin{equation}\label{10}
\begin{array}{l}
 {T_{\rm{CEI}}} =-4\pi^{3} V_e^{ - 2}\sum\limits_{q_1 q_2} \sum\limits_{s_1 s_2} p_1 [(u_{p_1}-s_1)\omega_1 + (u_{p_2}-s_2)\omega_2]\aleph_{s_1 s_2}(\zeta_2)\Phi_2(\textbf{p}_2)  \\
 ~~~~~~~~ \times I^{'}(\textbf{\textbf{P}}) \aleph_{d_1 d_2}(\zeta_1-\zeta_1^{'}) p_1^{'}[(u_{p_1}-q_1)\omega_1 + (u_{p_2}-q_2)\omega_2] \aleph_{q_1 q_2}(\zeta_1^{'})\Phi_1(\textbf{p}_1^{'}),   \\
 \end{array}
\end{equation}	
where
\begin{equation}\label{11}
\begin{array}{l}
 {I^{'}(\textbf{P})} =\int \int d\textbf{r}_1 d\textbf{r}_2 \exp[-i(\textbf{p}_1-\textbf{p}_1^{'})\cdot \textbf{r}_1]      \\
 ~~~~~~~~~~~~ \times \Phi_2^{*}(\textbf{r}_2) U_{12} \Phi_1(\textbf{r}_2). \\
 \end{array}
\end{equation}
Here, $d_1=(n_1-q_1)-(k_1+s_1)$ and $d_2=(n_2-q_2)-(k_2+s_2)$ are the photon
numbers absorbed from the first and second laser fields in the LACE process,
$\textbf{p}_1^{'}$ is the final momentum of the first ionized electron $e_1$
in the ATI1 process, $\Phi_2(\textbf{p}_2)$ is the wavefunction of the first
excited state of He$^+$ ion in the momentum space, and
$\Phi_1(\textbf{p}_1^{'})$ is approximated by the hydrogen-like wave
function with $I_{p_1}$ in the momentum space.

\section{NUMERICAL RESULTS}

\begin{flushleft}
  \textbf{1.~The momentum spectrum and channel analysis}
\end{flushleft}

We now consider NSDI process of a helium atom exposed to the IR and XUV
two-color laser fields, where the intensities of IR and XUV laser fields are
chosen as $I_1=1.0\times10^{12}$~W/cm$^{2}$ and
$I_2=5.0\times10^{13}$~W/cm$^{2}$, and the frequency of IR lase field is
$\omega_1=1.165$~eV. The polarization directions of the IR and XUV laser
fields are the same, and the initial phases of both laser fields are set to
zero for simplicity.

Figures~\ref{fig1}(a)-(d) show the NSDI momentum spectra of two ionized
electrons parallel to the laser polarization directions through the CEI
mechanism for different frequencies of XUV laser field (a)
$\omega_2=75\omega_1$, (b) $\omega_2=47\omega_1$, (c) $\omega_2=19\omega_1$
and (d) $\omega_2=13\omega_1$. For comparison, Figs.~\ref{fig1}(e)-(h) show
the momentum spectra through the CI mechanism under the same laser field
conditions, where the calculation formula is shown in Ref.~\cite{Jin1,Jin2}.
One can see that, as the energy of XUV photon decreases, the NSDI
probability of CEI mechanism is gradually higher than the corresponding
probability of CI mechanism. These results show that: if the energy of XUV
photon is small, the ionized electron cannot obtain enough energy by
absorbing XUV photons in the ATI1 process to make the bound electron ionized
directly by collision, while this bound electron can be easily pumped to the
excited state of He$^+$ ion by the laser-assisted collision and then is
ionized from the excited state by two-color laser fields at last. Hence the
CEI mechanism makes a dominant contribution to the NSDI process when the
frequency of XUV laser field is low. For example, in the case of
$\omega_2=19\omega_1$, the NSDI probability of CEI mechanism is larger than
that of the CI mechanism by about five orders of magnitude.

Figure~\ref{fig21} shows the NSDI momentum spectra through the CEI [(a)-(c)]
and CI [(d)-(f)] mechanisms for different laser intensities
$I_1=1.0\times10^{11}$~W/cm$^{2}$, $I_2=5.0\times10^{13}$~W/cm$^{2}$
[(a),(d)], $I_1=1.0\times10^{12}$~W/cm$^{2}$, $I_2=5.0\times10^{13}$~W/cm$^{2}$
[(b),(e)] and $I_1=1.0\times10^{12}$~W/cm$^{2}$,
$I_2=5.0\times10^{12}$~W/cm$^{2}$ [(c),(f)], in the case of
$\omega_2=19\omega_1$. One can see that the CEI mechanism still makes a
major contribution to the NSDI process for different laser intensities.
Furthermore, by comparing Fig.~\ref{fig21}(b) with Figs.~\ref{fig21}(a) and
(c), it is found that the XUV laser field determines the NSDI probability
and the IR laser field plays an important role in forming the structure of
the momentum spectrum, which is consistent with the situation under the CI
mechanism, as shown in Fig.~\ref{fig21}(e) and Figs.~\ref{fig21}(d) and (f). The
momentum spectra caused by CI mechanism have been investigated in the
previous paper~\cite{Jin1,Jin2}. Therefore, we focus on the NSDI process
dominated by the CEI mechanism in this paper.

Now we take the NSDI momentum in Fig.~\ref{fig1}(c) as example for analyzing
the CEI mechanism of NSDI process. One can firstly find that the momentum
distributions are the same in all the four quadrants, which is quite
different from that for the case of CI mechanism as shown in
Fig.~\ref{fig1}(g). This can be easily understood as follows: in the CEI
process, the first electron is ionized and it excites the second electron by
collision, then the second electron is ionized by the IR and XUV laser
fields from the excited state afterwards. As a result, the final momentum of
the first ionized electron is completely independent to that of the second
electron. Additionally, the two ionized electrons are indistinguishable, so
the interference patterns hold upon the exchange of the roles of the two
electrons. Therefore the CEI mechanism can lead to the same momentum
distribution in the four quadrants. Moreover, one can find that the NSDI
momentum spectrum can be divided into two regions consisting of the high and
low ionization probabilities, where the momentum spectrum shows complex
patterns. In the following, we will focus on analysis of the structure of
these momentum spectrum.

Based on the previous work~\cite{Hu2013,Zhang2013,Liu2015,Jin1,Jin2}, we
know that the XUV laser field has a decisive effect on the ionization
process in IR and XUV two-color laser fields. Figure~\ref{fig2} shows the
momentum spectra with the atom absorbing three (a), four (b), five (c) and
six (d) XUV photons in the NSDI process. One can see that the NSDI momentum
spectrum in Fig.~\ref{fig1}(c) is attributed to the processes of the atom
absorbing four and five XUV photons, as shown in Figs.~\ref{fig2}(b)
and~\ref{fig2}(c). This is because that, the double-ionization threshold is
so high that the atom has to absorb at least four XUV photons to make the
NSDI process happen. Furthermore, the processes of the atom absorbing four
and five XUV photons separately correspond to the two regions of the high
and low ionization probabilities, which indicates that the ionization
probability rapidly decreases with the increase of the number of the XUV
photons absorbed in the NSDI process.

As we mentioned in section II, we treat the NSDI as a three-step process:
the first electron is ionized in the ATI1 process, then collides with the
bound electron and makes it jump from the ground state to the first excited
state of He$^+$ ion in the LACE process, and finally the second electron is
ionized in the ATI2 process in the IR and XUV laser fields. In all the three
steps, the XUV photons may be absorbed by the electrons. Hence we now define
the channel as $(q_2,d_2,s_2)$, where $q_2$, $d_2$ and $s_2$ are the numbers
of XUV photons absorbed by electrons in the ATI1, LACE and ATI2 processes,
respectively. Firstly, we consider the region of high ionization probability
in the NSDI spectrum, where $q_2+d_2+s_2=4$. Figures~\ref{fig3}(a)-(c)
present the channel contributions of the NSDI momentum spectra for channels
(1,2,1) (a), (2,1,1) (b), and (3,0,1) (c). The contributions of the channels
(1,0,3), (1,1,2), (2,0,2) and the channels  with the atom absorbing zero XUV
photon in the ATI (including ATI1 or ATI2) process are not shown because of
their ignorable contributions to the NSDI process. From
Figs.~\ref{fig3}(a)-(c), one can clearly see that the region of the high
ionization probability is almost attributed to the contributions of channels
(2,1,1) and (3,0,1), where the contribution of channel (2,1,1) is the
greatest. Here we notice that, the final momentum of the first ionized
electron depends on both the first ionization potential $I_{p_1}$ of helium
and also the energy difference $\Delta E$ between the ground state  and the
first excited state of He$^+$ ion in the LACE process. Therefore, the first
electron at least absorbs three XUV photons in the ATI1 and LACE processes
to overcome these energies. Since $3\omega_2-I_{p_1}-\Delta E\approx 0$, the
final momentum of the first electron for channels (2,1,1) and (3,0,1) is
close to zero. On the other hand, the final momentum of the second ionized
electron only depends on the ionization potential $I_{p_{12}}$ of the first
excited state of He$^+$ ion, so the electron absorbs one XUV photon, which
is much larger than the ionization threshold, resulting in that the final
momentum of the second ionized electron is much greater than zero, as shown
in Figs.~\ref{fig3}(b) and \ref{fig3}(c).

Based on our theory, the NSDI caused by the CEI mechanism is due to the
laser-assisted collision-excitation process; hence the direction of the
first ionized electron's momentum may be changed before and after the
collision. Particularly, if the angle between the ionized electron's momenta
before and after the collision is smaller than $\pi /2$, we call this
collision a forward collision; otherwise, if the angle is larger than $\pi
/2$, we call it a backward collision.

Figures~\ref{fig3}(d)-(f) and (g)-(i) show the contributions of forward and
backward collisions for channels (1,2,1) [(d) and (g)], (2,1,1) [(e) and
(h)] and (3,0,1) [(f) and (i)], respectively. For channel (1,2,1), where the
first electron absorbs only one XUV photon in the ATI1 process, one can see
that the contribution of the backward collision is comparable with that of
the forward collision, while the momentum distribution of backward collision
is slightly broader than that of forward collision. For channels (2,1,1) and
(3,0,1), where the first electron carries high energy after it is ionized by
absorbing two or three XUV photons in the ATI1 process, the forward
collisions make a major contribution to the NSDI.

Similarly, Figures~\ref{fig4}(a)-(e) present the channel contributions of
the atom absorbing five XUV photons, and the other channels are not shown
due to little contributions. Here there are only five channels making
contributions to the region of the low ionization probability in the NSDI
process, where the contribution of the channel (3,1,1) is the biggest.
Furthermore, one can find that channels (2,2,1), (3,1,1) and (4,0,1)
illustrate similar interference patterns, while channels (2,1,2) and (3,0,2)
show similar patterns. By comparing Figs.~\ref{fig3}(a)-(c) with
Figs.~\ref{fig4}(a)-(e), one can find that these interference patterns
depend on the XUV photons absorbed in the ATI1+LACE processes and in the
ATI2 process. In order to analyze the contributions of forward and backward
collisions to NSDI process, figures~\ref{fig4}(f)-(j) and (k)-(o) show the
NSDI momentum spectra of the forward and backward collisions. One can see
that the forward collisions make major contributions to the NSDI process,
especially for the channels (3,0,2), (3,1,1) and (4,0,1). This result
further tells us that, if the ionized electron carries enough energy to
excite the bound electron, the forward collision dominates the NSDI process.

In order to further understand the above observation, we define subchannel
as $(q_2|q_1,d_2,s_2)$ within the channel $(q_2,d_2,s_2)$, where $q_1$ is
the number of the IR photons that the first electron absorbs ($q_1>0$) or
emits ($q_1<0$) in the ATI1 process. Now the LACE can be treated as the
process that the ionized electron with a certain value of $p'_1$ collides
with the parent ion, where
$p'_1=\sqrt{2[q_2\omega_2+q_1\omega_1-I_{p_{1}}-u_{p_1}\omega_1-u_{p_2}\omega_2]}$.
In the following, we will make a detailed discussion about the contributions
of forward and backward collisions for different values of $p'_1$.

Taking the subchannels of (2$|$-2,1,1), (2$|$0,1,1), (2$|$2,1,1) and
(2$|$5,1,1) within the channel (2,1,1) as examples, we present the NSDI
momentum spectra of these subchannels in Figs.~\ref{fig5}(a)-(d). One can
see that the momentum of one of two ionized electrons increases with the
increase of IR photons absorbed by an electron in the ATI1 process.
Furthermore, we present subchannel contributions of the forward collision in
Figs.~\ref{fig5}(e)-(h) and the backward collision in
Figs.~\ref{fig5}(i)-(l) for (2$|$-2,1,1) [(e) and (i)], (2$|$0,1,1) [(f) and
(j)], (2$|$2,1,1) [(g) and (k)] and (2$|$5,1,1) [(h) and (l)]. From
Fig.~\ref{fig5}, one can see that the forward collision provides a major
contribution to the NSDI, while the backward collision provides a broad
momentum distribution with a lower probability to the NSDI. This result
indicates that the electrons absorb more IR photons in the backward
collision than that in the forward collision, since the more IR photons the
electron absorbs, the lower NSDI probability is and the larger value that
the momentum presents.

Similarly, taking the subchannels of (3$|$-5,1,1), (3$|$-2,1,1), (3$|$1,1,1)
and (3$|$5,1,1) within the channel (3,1,1) as examples, we present the NSDI
momentum spectra of these subchannels in Figs.~\ref{fig6}(a)-(d) and the
subchannel contributions of the forward collision in
Figs.~\ref{fig6}(e)-(h), while the backward collisions are not shown because
of the little contribution to the corresponding subchannel. From
Figs.~\ref{fig6}(a) to \ref{fig6}(d), it is found that the momentum of the
first ionized electron increases with the increase of the number of the IR
photons that it absorbs in the ATI1 process.  In Fig.~\ref{fig6}, one can
see that the forward collision dominates the contribution to the NSDI.
Furthermore, by comparing Fig.~\ref{fig4}(d) with Figs.~\ref{fig6}(a)-(d),
one can find that the interference pattern in Fig.~\ref{fig4}(d) is
attributed to the coherent summation of these subchannel contributions.
Moreover, comparing Fig.~\ref{fig6} with Fig.~\ref{fig5}, it is found that,
for a given $q_1$, the relative contribution of forward collision to
backward collision for (3$|q_1$,1,1) is much larger than that for
(2$|q_1$,1,1). This indicates the relative contribution of forward to
backward collisions depends largely on the energy carried by the first
ionized electron after the ATI1 process.

\begin{flushleft}
  \textbf{2.~The saddle-point approximation and energy conservation}
\end{flushleft}

We now further explain why some channels illustrate similar interference
patterns, and why the relative contribution of forward to backward
collisions varies with the energy carried by the ionized electron in the
ATI1 process, as shown in Figs.~\ref{fig3}-\ref{fig4}. We employ the
saddle-point approximation to analyze the NSDI process. As mentioned above,
the NSDI can be treated as a three-step process. Firstly, we will consider
the first-step process, i.e., the first electron is ionized in the ATI1
process. We may analyze this process through the Bessel function
$\aleph_{q_1q_2}(\zeta_1')$ in Eq.~(\ref{10}). In our calculation, the
Bessel function can be simplified as~\cite{Zhang2013}
\begin{equation}\label{12}
 \aleph_{q_1q_2}(\zeta_1')\approx J_{-q_1}(\zeta_{11}',\zeta_{13}')J_{-q_2}(\zeta_{12}'),
\end{equation}
where
\begin{equation}\label{121}
 J_{-q_1}(\zeta_{11}',\zeta_{13}')=\sum\limits_{q_3}J_{-q_1+2q_3}(\zeta_{11}')J_{-q_3}(\zeta_{13}'),
\end{equation}
and
\begin{equation}\label{13}
\begin{gathered}
  {\zeta_{11}'}=2\sqrt {\frac{{{u_{p_1}}}}{{{\omega_1}}}} {{\mathbf{\text{{p}}}}_1'}\cos \theta_1', \hfill \\
  {\zeta_{12}'}=2\sqrt {\frac{{{u_{p_2}}}}{{{\omega_2}}}} {{\mathbf{\text{p}}}_1'} \cos \theta_1', \hfill \\
  {\zeta_{13}'} = \frac{1}{2}u_{p_1}, \hfill \\
\end{gathered}
\end{equation}
with $\theta_1'$ being the angle between the momentum of the first ionized
electron and polarization directions of the two laser fields in the ATI1
process.

For a certain $q_2$, the Bessel function $J_{-q_1}(\zeta_{11}',\zeta_{13}')$
in Eq.~(\ref{12}) can be written as~\cite{Wang2007,Guo2009}
\begin{equation}\label{14}
 J_{-q_1}(\zeta_{11}',\zeta_{13}')=\frac{1}{T_1}\int_{-T_1/2} ^{T_1/2} dt\exp{\{i[\zeta_{11}'\sin(\omega_1t)+\zeta_{13}'\sin(2\omega_1t)+q_1\omega_1t]\}},
\end{equation}
where $T_1=2\pi/\omega_1$. On the other hand, the classical action of an
electron is~\cite{Jin2}
\begin{equation}\label{15}
\begin{array}{l}
S_{cl}(t,\textbf{p}) =  \frac{1}{2}\int_0 ^t dt^{'} [\textbf{p}+\textbf{A}_{cl}(t^{'})]^2 \\
 ~~~~~~~~~~~=(\frac{1}{2}\textbf{p}^2+U_{p_1})t+2\sqrt {\frac{{{u_{p}}}_1}{{{\omega_1}}}} \sin(\omega_1 t) {{\mathbf{\text{\textbf{p}}}}}\cdot {{\widehat{{\mathbf{\epsilon}}}}_1}+\frac{1}{2}u_{p_1} \sin(2\omega_1 t),\\
 \end{array}
\end{equation}	
where $U_{p_1}$ is the ponderomotive energy from the IR laser field with the
vector potential $\textbf{A}_{cl}(t)$. By using Eq.~(\ref{15}) and the
energy conservation $E_i=E_m$, Eq.~(\ref{14}) becomes
\begin{equation}\label{16}
 J_{-q_1}(\zeta_{11}',\zeta_{13}')=\frac{1}{T_1}\int_{-T_1/2} ^{T_1/2} dt e^{i[f_{\textmd{ATI1}}(t)]},    \\
\end{equation}
where $f_{\textmd{ATI1}}(t)=S_{cl}(t,\textbf{p}_1')-(q_2
\omega_2-I_{p_{1}})t$. By applying the saddle-point approximation, the
saddle-point $t_1'$ satisfies $f'_{\textmd{ATI1}}(t)|_{t=t_1'}=0$, leading
to the energy conservation relationship in the ATI1 process
\begin{equation}\label{17}
 \frac{[\textbf{p}_1'+\textbf{A}_{cl}(t_1')]^{2}}{2}=q_2\omega_2-I_{p_{1}},
\end{equation}
 where $q_2$ is the number of XUV photons absorbed by the first ionized electron in the ATI1 process.

Secondly, we analyze the LACE process based on the properties of the Bessel
function $\aleph_{d_1 d_2}(\zeta_1-\zeta_1^{'})$ in Eq.~(\ref{10}). Under
the calculation condition that we use in this paper, this Bessel function
can be written as
\begin{equation}\label{18}
\aleph_{d_1 d_2}(\zeta_1-\zeta_1^{'}) = J_{-d_1}(\zeta_{11}-\zeta_{11}^{'})J_{-d_2}(\zeta_{12}-\zeta_{12}^{'}),
\end{equation}
where
\begin{equation}\label{19}
\begin{gathered}
{\zeta_{11}}=2\sqrt {\frac{{{u_{p_1}}}}{{{\omega_1}}}} {{\mathbf{\text{{p}}}}_1} \cos \theta_1 , \\
{\zeta_{12}}=2\sqrt {\frac{{{u_{p_2}}}}{{{\omega_2}}}} {{\mathbf{\text{p}}}_1} \cos \theta_1 . \\
\end{gathered}
\end{equation}
Here, $\theta_1$ is set to zero for the final momentum of the first ionized
electron along the polarization direction of the two laser fields.

The Bessel function $J_{-d_1}(\zeta_{11}-\zeta_{11}^{'})$ in Eq.~(\ref{18})
can also be expressed as an integral form~\cite{Wang2007,Guo2009}
\begin{equation}\label{20}
 J_{-d_1}(\zeta_{11}-\zeta_{11}^{'})=\frac{1}{T_1}\int_{-T_1/2} ^{T_1/2} dt\exp{\{i[(\zeta_{11}-\zeta_{11}^{'})\sin(\omega_1t)+d_1\omega_1t]\}}.
\end{equation}
By using Eq.~(\ref{15}) and the energy conservation $E_m=E_l$ in the LACE
process, Eq.~(\ref{20}) can be rewritten as
\begin{equation}\label{21}
 J_{-d_1}(\zeta_{11}-\zeta_{11}^{'})=\frac{1}{T_1}\int_{-T_1/2} ^{T_1/2} dt e^{i[f_{\textmd{LACE}}(t)]}, \\
\end{equation}
where $f_{\textmd{LACE}}(t)=\Delta S_{cl}-(d_2 \omega_2-\Delta E)t$ with
$\Delta S_{cl}= S_{cl}(t,\textbf{p}_1)-S_{cl}(t,\textbf{p}_1')$ and $\Delta
E=I_{p_{2}}-I_{p_{12}}$. Under the saddle-point approximation, this Bessel
function will further become
\begin{equation}\label{22}
\begin{array}{l}
J_{-d_1}(\zeta_{11}-\zeta_{11}^{'})\approx \sum\limits_{t_1}\frac{1}{\sqrt{\pi}[d^2_1-(\zeta_{11}-\zeta_{11}^{'})^2]^{1/4}}  \\
~~~~~~~~~~~~~~~~~~~~\times \exp\{-i[(\zeta_{11}^{'}-\zeta_{11})\sin(\omega_1t_1)-d_1\omega_1t_1]\},
 \end{array}
\end{equation}
where the saddle-point $t_1$ satisfies
\begin{equation}\label{23}
\frac{1}{2}{[\textbf{p}_1+\textbf{A}_{cl}(t_1)]}^2=\frac{1}{2}{[\textbf{p}'_1+\textbf{A}_{cl}(t_1)]}^2+(d_2\omega_2-\Delta E),
\end{equation}
which is the energy conservation equation of the first ionized electron in
the LACE process. It shows that the collision is an inelastic process with
absorbing $d_2$ XUV photons and overcoming the energy $\Delta E$. Then, we
obtain the following equation according to Eq.~(\ref{17}) and Eq.~(\ref{23})
\begin{equation}\label{24}
\frac{1}{2}{[\textbf{p}_1+\textbf{A}_{cl}(t_1)]}^2=(q_2+d_2)\omega_2-I_p',
\end{equation}
where $I_p'=I_{p_1}+I_{p_2}-I_{p_{12}}$. This equation indicates that the
final momentum spectrum of the first electron depends on the total XUV
photons $(q_2+d_2)$ that it absorbs in the ATI1 and LACE processes. This is
the reason why the structures of the momentum spectra of the subchannels
(2,1,1) and (3,0,1) are the same.

Lastly, we will consider the third-step process, i.e., the second electron
is ionized from the first excited state of He$^+$ ion by the IR+XUV
two-color laser fields. Similarly, by analyzing the Bessel function
$\aleph_{s_1s_2}(\zeta_2)$ in Eq.~(\ref{10}) under the saddle-point
approximation,  we can also obtain the energy conservation equation
satisfied by the second electron in the ATI2 process
\begin{equation}\label{25}
 \frac{[\textbf{p}_2+\textbf{A}_{cl}(t_2)]^{2}}{2}=s_2\omega_2-I_{p_{12}},
\end{equation}
where $s_2$ is the number of XUV photons absorbed by the second ionized
electron in the ATI2 process. In the above, the $t_2$ satisfies the equation
$f'_{\textmd{ATI2}}(t)|_{t=t_2}=0$ with
$f_{\textmd{ATI2}}(t)=S_{cl}(t,\textbf{p}_2)-(s_2 \omega_2-I_{p_{12}})t$.

To explain the reason of similar interference patterns of different
channels, Fig.~\ref{fig7} presents comparisons of ATI spectra and NSDI
spectra for channel (a) (2,1,1), (b) (3,1,1) and (c) (2,1,2). To understand
the structure, we show the ATI spectra based on Eq.~(\ref{901}) and
Eq.~(\ref{902}) in Figs.~\ref{fig7}(d) and \ref{fig7}(e), respectively. In
Figs.~\ref{fig7}(d), the $I_{p_1}$ is replaced by $I'_p$ in Eq.~(\ref{901}).
From Figs.~\ref{fig7}(d) and \ref{fig7}(e), one can see that both the ATI
spectra present steplike structures~\cite{Zhang2013}. According to
Eq.~(\ref{24}), one can find that the first and second steplike structures
in Fig.~\ref{fig7}(d) are attributed to the processes of atom absorbing
three and four XUV photons under the IR laser field in the ATI1+LACE
processes, respectively. Similarly, the first and second steplike structures
in Fig.~\ref{fig7}(e) are due to the first-excited electron of He$^+$ ion
absorbing one and two XUV photons under the IR laser field in the ATI2
process, respectively. By mapping the ATI spectrum of Fig.~\ref{fig7}(d) (
Fig.~\ref{fig7}(e)) on the spectra of Figs.~\ref{fig7}(a) to \ref{fig7}(c)
with dash (solid) lines, one can see that the outline of interference
patterns of these NSDI spectra can be well reproduced by the steplike
structures in the ATI spectra, which indicates that the NSDI process caused
by the CEI mechanism can be decoupled into two ATI processes of the
ATI1+LACE and ATI2. It is the reason why Figs.~\ref{fig3}(b) and
\ref{fig3}(c) illustrate the similar interference patterns, as well as
Figs.~\ref{fig4}(b), \ref{fig4}(d) and \ref{fig4}(e) show the similar
patterns, and also Figs.~\ref{fig4}(a) and \ref{fig4}(c) exhibit same
patterns.

On the other hand, in order to further understand the contributions of
forward and backward collisions to the NSDI, as shown in
Figs.~\ref{fig3}-\ref{fig4}, we focus on the LACE process by analyzing the
Bessel function $\aleph_{d_1 d_2}(\zeta_1-\zeta_1^{'})$ in Eq.~(\ref{10}).
One can find that the Bessel function determines the probability of the LACE
process and can also be expressed as $\aleph_{d_1 d_2}(\zeta_1-\zeta_1^{'})
\propto J_{-d_1}(\zeta_{11}-\zeta_{11}^{'})$ for the weighting factor
$J_{-d_2}(\zeta_{12}-\zeta_{12}^{'})$. Furthermore, the Bessel function
$J_{-d_1}(\zeta_{11}-\zeta_{11}^{'})$  is proportional to $
\frac{1}{f^{1/4}}$, where $f=d^2_1-
\frac{4u_{p_1}}{\omega_1}({{\mathbf{\text{{p}}}}_1^2}+{{\mathbf{\text{{p}}}'}_1^2}\cos^2\theta_1'-2{{\mathbf{\text{{p}}}}_1}{{\mathbf{\text{{p}}}}_1'}\cos
\theta_1')$. Hence, the smaller the value of $f$ is, the larger the value
that $J_{-d_1}(\zeta_{11}-\zeta_{11}^{'})$ can have. If $f$ infinitely goes
closer to zero and $d_1 \geq 0$, we get
\begin{equation}\label{26}
d_1=2\sqrt{\frac{u_{p_1}}{\omega_1}({{\mathbf{\text{{p}}}}_1^2}+{{\mathbf{\text{{p}}}'}_1^2}\cos^2\theta_1'-2{{\mathbf{\text{{p}}}}_1}{{\mathbf{\text{{p}}}}_1'}\cos \theta_1') }.
\end{equation}
In the above, one can see that, for the maximum value of
$J_{-d_1}(\zeta_{11}-\zeta_{11}^{'})$, $d_1$ is determined by $p_1$ and
$\theta'_1$ for a given value $p'_1$.

In the LACE process, the IR laser field may be treated as a classical field
in the saddle-point approximation. Figure~\ref{fig8} shows IR photon numbers
$d_1$ in the LACE process as a function of $p_1$, which is the momentum
value of the ionized electron after collision, and $\theta'_1$ with the
momentum value before collision $p'_1$ being (a) 0.26~a.u., (b) $1.21$~a.u.
and (c) $2.17$~a.u., where the region of $\theta'_1<\pi/2$ in each graph
represents forward collision and the region of $\theta'_1>\pi/2$ represents
backward collision. One can see that, for backward collision, the number of
the IR photons absorbed by the first electron increases as the value of
$p'_1$ increases. We have noticed that, although the atom absorbs a few IR
photons (less than 12), the total energy of these IR photons is not enough
to excite the bound electron, hence the source of the energy $\Delta E$ to
excite the bound electron comes not only from the energy transfer between
the two electrons and absorbing IR photons, but also from absorbing XUV
photons during the collision. This result is consistent with the calculated
result of Hu~\cite{Hu2013}. Furthermore, for the case of absorbing certain
number of XUV photons, the probability of the LACE process is determined by
the number of IR photons that electron absorbs. From Fig.~\ref{fig8}, it is
found that: (i) when the value of $p'_1$ is small, the numbers of the IR
photons absorbed in both the forward and backward collisions are about equal
as shown in Fig.~\ref{fig8} (a), and hence the forward and backward
collisions make a comparable contribution to the NSDI, as shown in
Figs.~\ref{fig3}(d) and \ref{fig3}(g); (ii) the number of the IR photons
absorbed in the forward collision is less than that in the backward
collision when the value of $p'_1$ becomes larger as shown in
Fig.~\ref{fig8} (b), which indicates that the forward collision makes a
major contribution to the NSDI, while the backward collision provides a
broader momentum distribution, as shown in Fig.~\ref{fig5}; (iii) when the
ionized electron carries enough energy to excite the bound electron by
collision without absorbing the XUV photons in the LACE process, the
electron absorbs much less IR photons in the forward collision than that it
does in backward collision as shown in Fig.~\ref{fig8} (c), thus the forward
collision plays a dominant role in the NSDI, as shown in Fig.~\ref{fig3}(f)
as well as Figs.~\ref{fig4}(c)and (e). To summarize the above analysis, with
increase of the value of $p'_1$, the source of the energy $\Delta E$ to
excite the bound electron changes from absorbing XUV photons to the pure
energy transfer between the two electrons in the LACE process, and hence the
forward collision gradually becomes dominant in the NSDI, as shown in
Figs.~\ref{fig3}-\ref{fig6}.

In contrast to the above results, the backward collision makes a dominant
contribution to the NSDI in the case of a monochromatic IR laser field,
which is shown in Fig.8 of Ref.~\cite{Wang2012}. Here we may notice that, in
a monochromatic IR laser field, the amplitude of the LACE process only
depends on Eq.~(\ref{20}), and hence the small value of $p'_1$ in the ATI1
process makes a crucial contribution to the NSDI. In order to explain the
difference, Fig.~\ref{fig9} shows a comparison of IR photon number $d_1$ in
the LACE process under the monochromatic IR (a) and IR+XUV two-color (b)
laser fields for $p'_1=0.26$~a.u.. The intensity and frequency of
monochromatic IR laser field in Fig.~\ref{fig9}(a) are
$I=2.2\times10^{14}$~W/cm$^{2}$ and $\omega=1.165$~eV. Figure~\ref{fig9}(b)
shows a part of Fig.~\ref{fig8}(a) for small value of $p_1$.  One can see
that, the electron absorbs more IR photons for the backward collision than
that for the forward collision in both cases, while the range of the number
of IR photons absorbed in the monochromatic IR laser field is much larger
than that in two-color laser fields. Furthermore, in Fig.~\ref{fig9}(a), one
can find that the energy obtained in the forward collision is so little that
the forward collision can not make the bound electron to be excited. On the
contrary, the electron can absorb lots of IR photons in the backward
collision, and hence the bound electron may have a chance to be excited in
the LACE process. As a result, the backward collision makes a crucial
contribution to the NSDI in the monochromatic IR laser field.

\section{CONCLUSIONS}
Based on the frequency-domain theory, we investigate the CEI mechanism of
NSDI process for a helium atom in the IR+XUV two-color laser fields. We find
that the NSDI probability of  the CEI mechanism is much larger than that of
the CI mechanism when the XUV photon energy is smaller than the ionization
threshold of He$^+$ ion. It shows that the NSDI momentum spectrum caused by
the CEI mechanism presents complex interference patterns and symmetrical
structures. With the help of channel analysis, we find that the momentum
spectrum is attributed to the interference between different channels. With
the channel number increases, the energy of the ionized electron increases
in the ATI1 process, and hence the source of the energy to excite the bound
electron changes from absorbing XUV photons to pure energy transfer between
two electrons during the LACE process. Furthermore, in order to explain the
interference pattern, we investigated the contributions to the NSDI from the
forward and backward collisions. We found that, if the energy of recolliding
electron is small, both the forward and backward collisions play comparable
roles in the NSDI, while the distribution of momentum spectrum is broader
for the backward collision than for the forward; on the other side, if the
the energy of recolliding electron is very large, the forward collision
plays a dominant role in the NSDI. Also, we employed the saddle-point
approximation to obtain the equation of energy conservation, and
reconstructed the formation of interference patterns of different channels
by the spectra of two ATI processes. In addition, we also explained the
reason why the backward collision plays a dominant role in the CEI mechanism
of NSDI under the monochromatic IR laser field.

\section*{ACKNOWLEDGMENTS}
This research was supported by the National Natural Science Foundation of
China under Grant Nos. 11474348, 61275128, 11334009 and 11425414. C. J. was
supported by the National Basic Research Program of China Grant No.
2013CB922201. Z.-C.Y. was supported by the Natural Sciences and Engineering
Research Council of Canada and by the Canadian computing facilities of
SHARCnet, and ACEnet.

\newpage{
\begin{figure}
\includegraphics[width=0.7\textwidth]{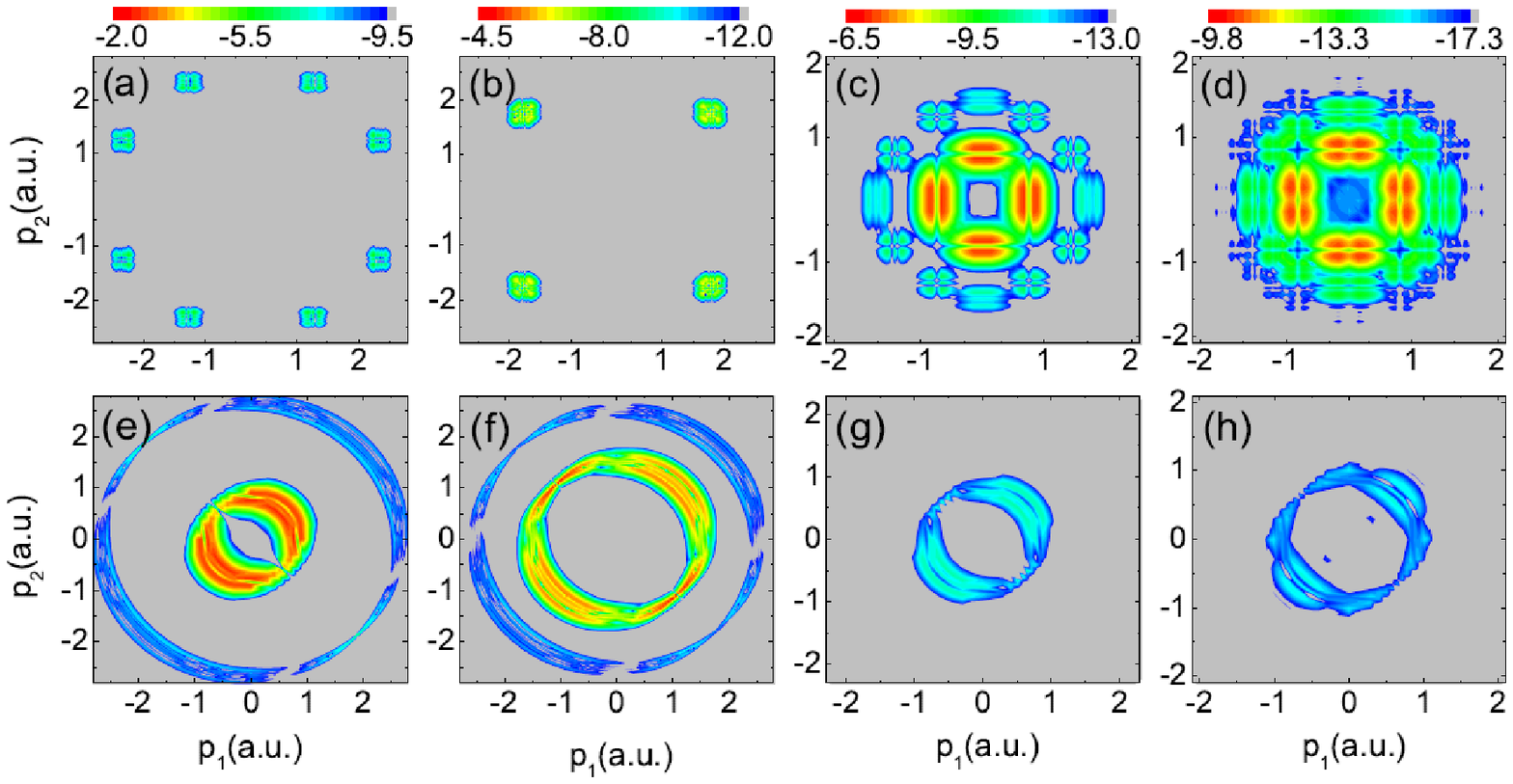}
\caption{(Color online). Momentum spectra of two ionized electrons along the laser polarization directions through [(a)-(d)] the collision-excitation ionization and [(e)-(h)] the collision-ionization processes for different laser frequencies $\omega_2=75\omega_1$ [(a),(e)], $\omega_2=47\omega_1$ [(b),(f)], $\omega_2=19\omega_1$ [(c),(g)] and $\omega_2=13\omega_1$ [(d),(h)], where the intensities of IR and XUV laser fields are $I_1=1.0\times10^{12}$~W/cm$^{2}$ and $I_2=5.0\times10^{13}$~W/cm$^{2}$, and the frequency of IR lase field is $\omega_1=1.165$~eV. In logarithmic scale.}
\label{fig1}
\end{figure}}

\newpage{
\begin{figure}
\includegraphics[width=0.45\textwidth]{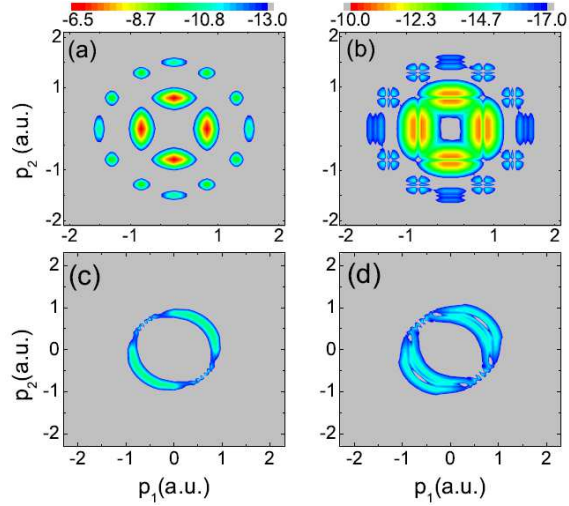}
\caption{(Color online). The NSDI momentum spectra through the CEI [(a)-(c)] and CI [(d)-(f)] mechanisms for different laser intensities $I_1=1.0\times10^{11}$~W/cm$^{2}$, $I_2=5.0\times10^{13}$~W/cm$^{2}$ [(a),(d)], $I_1=1.0\times10^{12}$~W/cm$^{2}$, $I_2=5.0\times10^{13}$~W/cm$^{2}$ [(b),(e)] and $I_1=1.0\times10^{12}$~W/cm$^{2}$, $I_2=5.0\times10^{12}$~W/cm$^{2}$ [(c),(f)], where the frequency of XUV laser field is $\omega_2=19\omega_1$. In logarithmic scale.}
\label{fig21}
\end{figure}}

\newpage{
\begin{figure}
\includegraphics[width=0.6\textwidth]{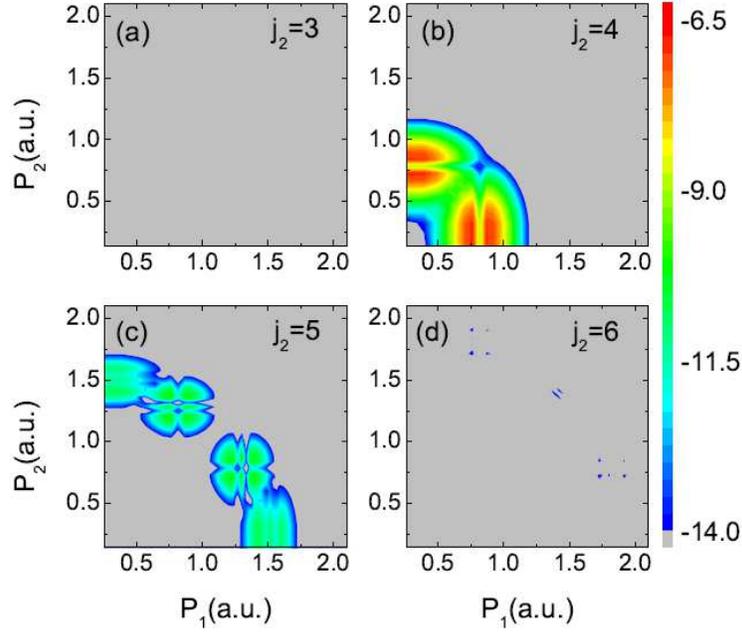}
\caption{(Color online). Momentum spectra of the atom totally absorbing (a) $ j_2$=3, (b) $ j_2$=4, (c) $ j_2$=5 and (d) $j_2$=6 XUV photons in the NSDI process. In logarithmic scale.}
\label{fig2}
\end{figure}}

\newpage{
\begin{figure}
\includegraphics[width=0.6\textwidth]{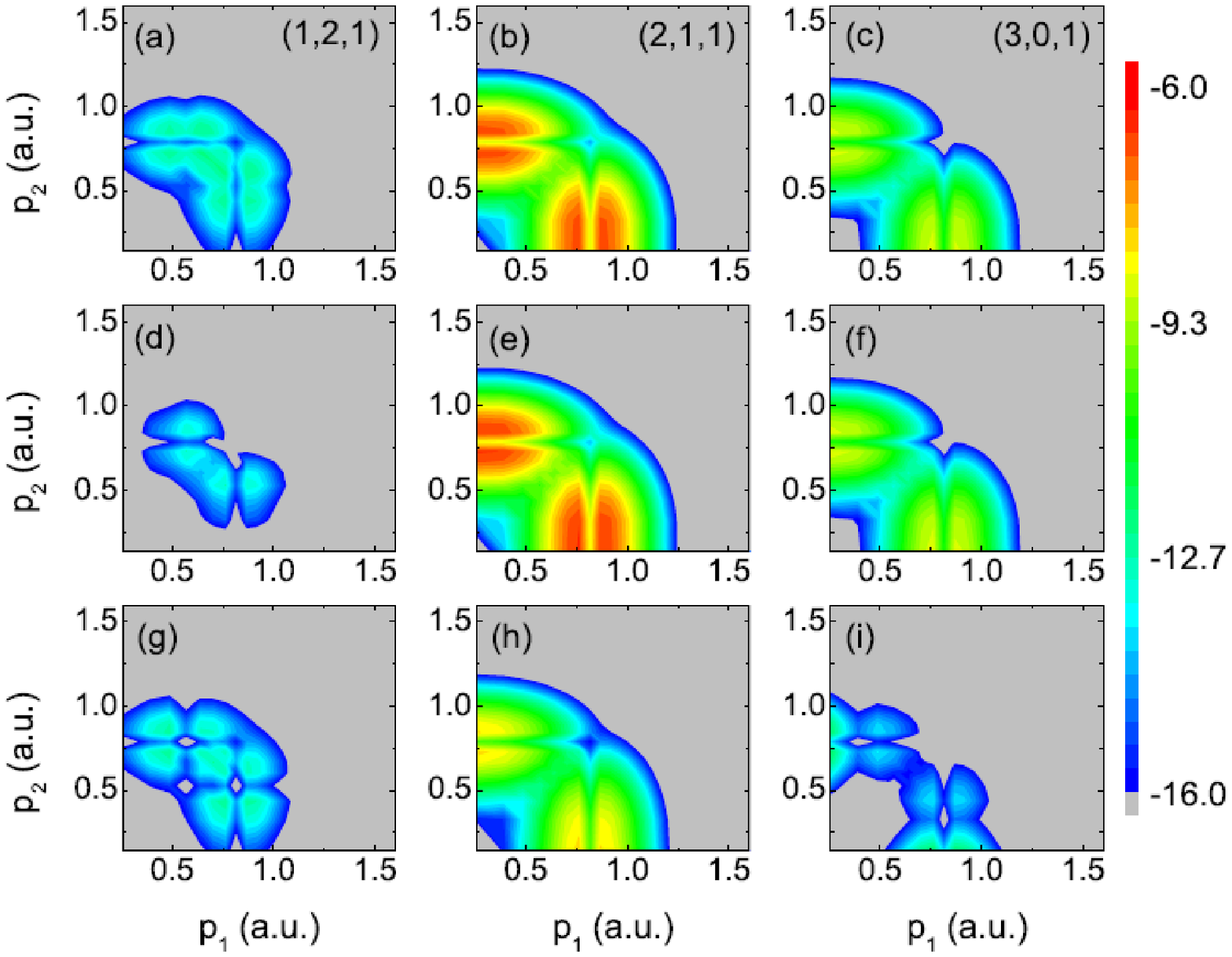}
\caption{(Color online). Channel contributions of the NSDI momentum spectra for channels (a) (1,2,1), (b) (2,1,1), (c) (3,0,1).  (d)-(f) and (d)-(f) present the contributions of forward and backward collisions for channels [(d), (g)] (1,2,1), [(e), (h)] (2,1,1) and [(f), (i)] (3,0,1), respectively. In logarithmic scale.}
\label{fig3}
\end{figure}}

\newpage{
\begin{figure}
\includegraphics[width=0.6\textwidth]{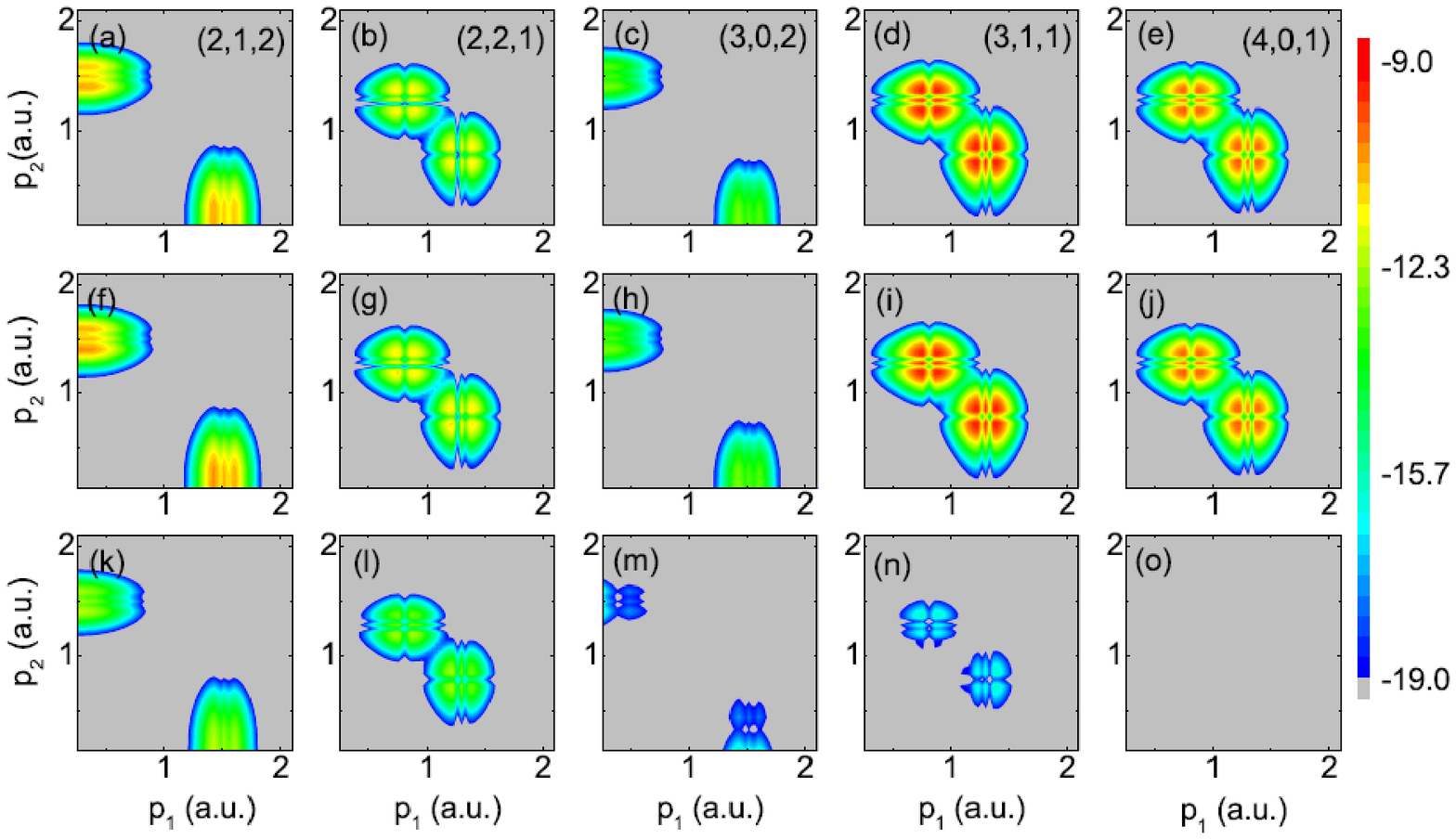}
\caption{(Color online). Channel contributions of the NSDI momentum spectra for channels (a) (2,1,2), (b) (2,2,1), (c) (3,0,2), (d) (3,1,1) and (e) (4,0,1). (f)-(j) and (d)-(f) present the contributions of forward and backward collisions for channels [(f), (k)] (2,1,2), [(g), (l)] (2,2,1), [(h), (m)] (3,0,2), [(i), (n)] (3,1,1) and [(j), (o)] (4,0,1), respectively. In logarithmic scale.}
\label{fig4}
\end{figure}}

\newpage{
\begin{figure}
\includegraphics[width=0.6\textwidth]{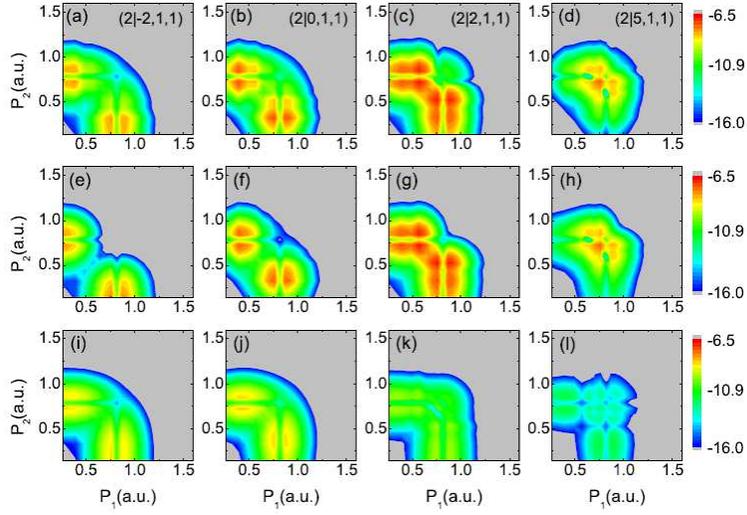}
\caption{(Color online).  Subchannel contributions of the NSDI momentum spectra for (a) (2$|$-2,1,1), (b) (2$|$0,1,1), (c) (2$|$2,1,1) and (d) (2$|$5,1,1). (e)-(h) and (i)-(l) present the NSDI momentum spectra of the forward and backward collisions for subchannels [(e), (i)] (2$|$-2,1,1), [(f), (j)] (2$|$0,1,1), [(g), (k)] (2$|$2,1,1) and [(h), (l)] (2$|$5,1,1), respectively. In logarithmic scale.}
\label{fig5}
\end{figure}}

\newpage{
\begin{figure}
\includegraphics[width=0.8\textwidth]{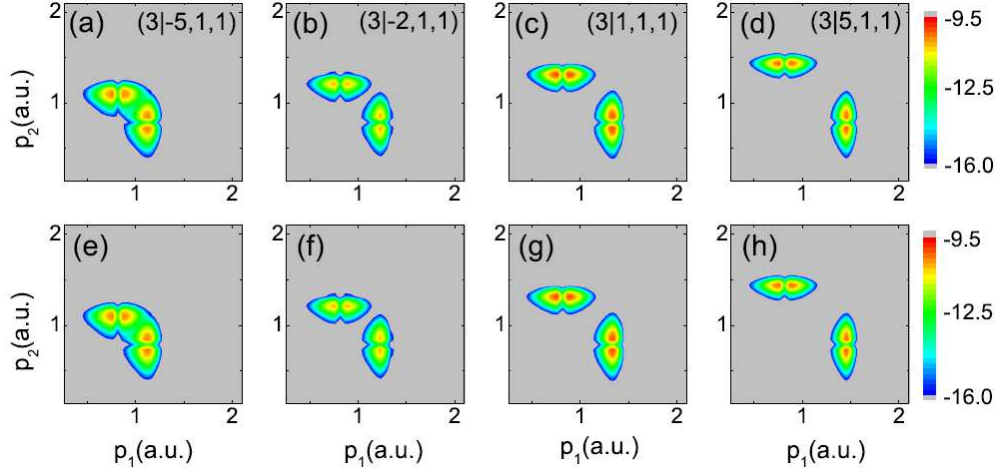}
\caption{(Color online). Subchannel contributions of the NSDI momentum spectra
for (a) (3$|$-5,1,1), (b) (3$|$-2,1,1), (c) (3$|$1,1,1) and (d) (3$|$5,1,1).
(e)-(h) present the NSDI momentum spectra of the forward collision for subchannel (e) (3$|$-5,1,1), (f) (3$|$-2,1,1), (g) (3$|$1,1,1) and (h) (3$|$5,1,1). In logarithmic scale.}
\label{fig6}
\end{figure}}

\newpage{
\begin{figure}
\includegraphics[width=0.55\textwidth]{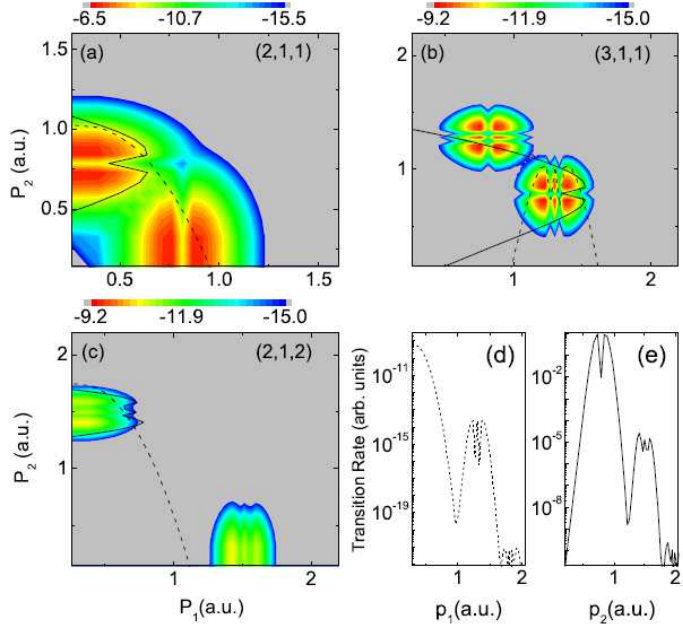}
\caption{(Color online). Comparisons of ATI spectra and NSDI spectra for channel (a) (2,1,1), (b) (3,1,1) and (c) (2,1,2).
The dash lines in (a), (b) and (c) represent the ATI spectrum based on Eq.~(\ref{901}) with $I_{p_1}$ replaced by $I'_p$, where this ATI spectrum is shown in (d);
and the solid lines in (a), (b) and (c) represent the ATI spectrum based on Eq.~(\ref{902}), where this ATI spectrum is shown (e).} \label{fig7}
\end{figure}}

\newpage{
\begin{figure}
\includegraphics[width=0.55\textwidth]{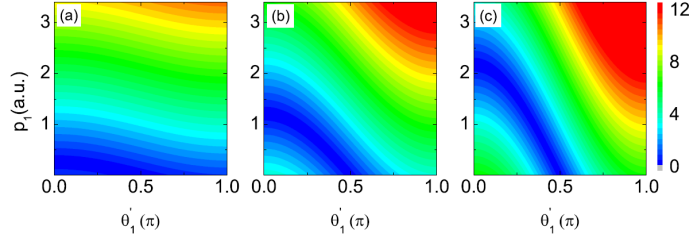}
\caption{(Color online). Number of atoms absorbing IR photons in the LACE process
for (a) $p'_1=0.26$~a.u., (b) $p'_1=1.21$~a.u. and (c) $p'_1=2.17$~a.u.,
where the direction of the momentum $\textbf{p}'_1$ varies from 0 to $\pi$,
and the momentum $\textbf{p}_1$ is along the IR laser polarization direction.
The region of $\theta'_1<\pi/2$ in each graph represents forward collision and the region of $\theta'_1>\pi/2$ in each graph represents backward collision.}
\label{fig8}
\end{figure}}

\newpage{
\begin{figure}
\includegraphics[width=0.55\textwidth]{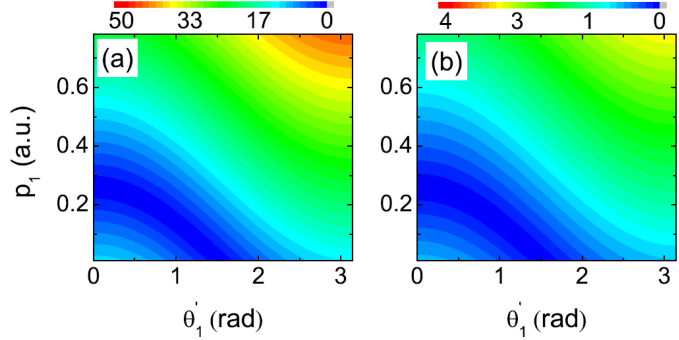}
\caption{(Color online). Comparisons of absorbing IR photons in the LACE process
under (a) the monochromatic IR and (b) IR+XUV two-color laser fields for $p'_1=0.26$~a.u..
The left is for the intensity and frequency of IR laser field being $I=2.2\times10^{14}$~W/cm$^{2}$
and $\omega=1.165$~eV, and the right is the part of Fig.~\ref{fig8}(a) for small value $p_1$.
The region of $\theta'_1<\pi/2$ in each graph represents forward collision and the region of $\theta'_1>\pi/2$ in each graph represents backward collision.}
\label{fig9}
\end{figure}}


\end{document}